\documentclass[a4paper,fleqn]{cas-dc}

\usepackage[authoryear,longnamesfirst]{natbib}
\usepackage{booktabs}
\usepackage{adjustbox}
\usepackage{array}
\usepackage{makecell}
\usepackage{float}
\usepackage{graphicx}
\usepackage{caption}
\usepackage{threeparttable}
\usepackage{cuted}

\def\tsc#1{\csdef{#1}{\textsc{\lowercase{#1}}\xspace}}
\tsc{WGM}
\tsc{QE}
\tsc{EP}
\tsc{PMS}
\tsc{BEC}
\tsc{DE}

\begin{document}
\let\WriteBookmarks\relax
\def\floatpagepagefraction{1}
\def\textpagefraction{.001}
\shorttitle{Consequences of high SMR operating costs}
\shortauthors{P. Rao et~al.}

\title [mode = title]{The consequences of high SMR operating costs in electricity markets}         



\author[1]{Pradyumna Rao}[type=editor,
                        orcid=0009-0005-8063-2271]
\cormark[2]
\ead{prady.rao@gmail.com}

\credit{Conceptualization of this study, Data curation, Methodology, Software, Writing - Original draft preparation}

\affiliation[1]{organization={Paul M. Rady Department of Mechanical Engineering and Renewable and Sustainable Energy Institute, University of Colorado-Boulder},
            addressline={1111 Engineering Drive}, 
            city={Boulder},
            postcode={80309}, 
            state={CO},
            country={United States}}

\author[2]{Daniel T. Kaffine}[
                        orcid=0000-0002-9556-7776]
\ead{daniel.kaffine@colorado.edu}
\ead[url]{https://www.colorado.edu/faculty/kaffine/}
\affiliation[2]{organization={Department of Economics and Renewable and Sustainable Energy Institute, University of Colorado Boulder},
            addressline={Economics Building, 261 UCB \#212}, 
            city={Boulder},
            postcode={80302}, 
            state={CO},
            country={United States}}
\credit{Conceptualization of this study, Writing - Review \& Editing, Supervision}



\affiliation[3]{organization={Department of Electrical, Computer, and Energy Engineering and Renewable and Sustainable Energy Institute, University of Colorado Boulder},
            addressline={4001 Discovery Dr. N321}, 
            city={Boulder},
            postcode={80309}, 
            state={CO},
            country={United States}}

\author[3]{Bri-Mathias Hodge}[orcid=0000-0001-8684-0534]
\cormark[1]
\fnmark[1,3]
\ead{brimathias.hodge@colorado.edu}
\ead[URL]{https://www.colorado.edu/faculty/hodge/}
\credit{Conceptualization of this study, Writing - Review \& Editing, Investigation, Supervision}

\cortext[cor1]{Corresponding author}
\cortext[cor2]{Principal corresponding author}

\begin{abstract}
As US power markets contend with growing demand for firm generation, the nuclear industry has offered Small Modular Reactors (SMRs). However, how these concepts would fare in a rapidly evolving power grid is unclear, given the paucity of operational examples. Current literature, informed by substantial cost escalations for traditional nuclear plants, focuses on the investment costs SMRs need to achieve for private investment feasibility. However, this work finds that the operating and marginal costs of SMRs are more critical to economic feasibility in market environments. This work dispatches SMRs using a flexible operations model, considering revenue from two main electric markets, capacity and wholesale energy markets, with and without policy support. Manufacturer advertised costs for investment and operating costs are used, with fuel costs calculated from manufacturer provided design parameters. Results indicate that SMRs are uneconomical primarily because investment cost reductions are offset by increased marginal costs. As such, an environment of prices and subsidies beyond historic norms are necessary to attract private investment at manufacturer advertised cost benchmarks. Current SMRs are as profitable as advanced estimates of the AP1000 traditional nuclear reactor, and if investment costs escalate at the average rate for nuclear projects, they are similar to Vogtle 3 \& 4. In projected future power markets, reductions in marginal cost may be more beneficial than those in investment costs.
\end{abstract}


\begin{highlights}
\item While SMRs are designed for lower investment costs, increased operating costs lead them to be uneconomical
\item SMRs at manufacturer advertised costs are unable to pay back in less than their lifetime in wholesale markets.
\item Subsidies and prices need to exceed historic highs for SMRs to be economically viable
\item Improvements in marginal costs may be more valuable than in investment costs for SMRs in future power grids
\end{highlights}


\begin{keywords}
Small Modular Reactors
\sep Nuclear Power
\sep Electricity Markets
\sep Tax Credits
\end{keywords}

\maketitle

\section{Introduction}\label{introsec}

Momentum for nuclear generation has been building over the past few years as demand for carbon-free firm generation grows. The restart of Three Mile Island by Microsoft Corporation \cite{constellation_crane_clean_energy_center}, Power Purchase Agreements by Amazon with Talen and Energy Northwest for traditional and Small Modular Reactors \cite{Amazon2023}, and Google's agreements with Kairos Power \cite{googlekairos} are examples of material commitments made towards building new nuclear power. These efforts follow declarations made by 25 countries at COP28 to triple nuclear capacity by 2050 \cite{DOE2023} to achieve global net-zero goals, such as climate benchmarks of limiting average temperature increase to \ensuremath{1.5^\circ\text{C}} above pre-industrial levels \cite{UN2023degree}. 

However, experiences building new nuclear plants in the United States and France over the past few decades have incurred substantial cost escalation, classified as ``negative learning by doing" \cite{grubler2010}. Additionally, long lead times, safety issues, permanent disposal of radioactive waste, and proliferation of potential nuclear weapon material \cite{ADAMANTIADES20095149, mit_nuclear_power_future_2003} have increased investor reticence towards new gigawatt scale nuclear power plants \cite{locatelli_2018_why_megaprojects}. Existing nuclear power plants have retired due to difficulty remaining competitive \cite{HARATYK2017150} in electricity markets, attributed to downward pressure on electricity prices from increased penetration of renewable energy \cite{lbnlelecprices}. 

Therefore, central to achieving these nuclear buildout targets are a new generation of nuclear reactors called Small Modular Reactors, commonly known as SMRs. Classified as reactors with power outputs $<$ 300 MWe, these prototypes are part of a concerted effort over the last two decades by the nuclear industry to address techno-economic concerns surrounding traditional gigawatt-scale reactors. SMRs have been touted as simpler, safer, and cheaper due to their modular designs \cite{vujic2012}. Additionally, load-following features of SMRs aim to improve integration with variable renewable power \cite{locatelliloadfollowing}, as well as enable generation or co-generation with industrial products such as hydrogen \cite{pham2022}, steam and thermal heat \cite{vanatta2024}. However, the private investment potential of SMRs and nuclear power for future power grids is a source of intensifying debate \cite{KIM2026105989, steigerwald2023, jnnolandieee, schlissel2024small, froese2020too}.

This work investigates the economic performance of SMRs under historical and future power scenarios, with and without additional policy support. Independent System Operator (ISO) Markets are defined by their two components - wholesale energy markets and capacity markets. Manufacturer-advertised SMR cost estimates should be interpreted with caution, as their veracity remains disputed \cite{steigerwald2023,Schlissel2023_IEEFA_NuScaleCostEstimates}. However, given the limited availability of SMR cost estimates validated through operating experience, this work uses these advertised values as inputs for simulating economic performance. The primary research question this work seeks to answer is: are SMR concepts economically viable from a private investment perspective? With over 70 SMR concepts under development globally \cite{IAEA2024SMR}, whether these concepts would be more economical and attractive as private investment vehicles is key to their prospects. To explore how SMRs would perform in future power grids, this paper primarily builds on two main works from the literature. 

The primary work this paper builds on is \cite{steigerwald2023}, who find after comparing manufacturer and production-theory based estimates, no SMR concept would be competitive or profitable. While comparing Levelized Cost of Electricity (LCOE) and Net Present Value (NPV) to alternate investment options, such as existing renewable technologies, was the focus of the work, a sub-finding showed low to negative NPVs using SMR manufacturer advertised estimates. Exploring causal factors of the finding on poor economic outcomes at SMR manufacturer costs is the primary motivation behind this work. Limiting factors of \cite{steigerwald2023} were their use of static assumptions on operational characteristics such as capacity factors and sampled constant lifetime prices. This fails to capture how evolving features of the grid may affect SMR investments \cite{kings_college}, from temporal price volatility \cite{iese_price_article} to their corresponding operational decisions, such as ramping in response to low or negative prices, which will play a significant role for thermal power feasibility in future low-carbon grids \cite{cycling_impacts_feasibility}. As such, in calculating feasibility metrics such as NPV and discounted payback periods, this work dispatches a flexible operations model based on literature methodologies \cite{alhadhrami2023} on hourly price profiles to endogenize temporal and operational impacts. This work also extends manufacturer advertised SMR cost parameters \cite{steigerwald2023} with technical design constraints and refueling outages to supplement the operational model.

The second key related work is \cite{HARATYK2017150}, who found that existing nuclear power plants remain uncompetitive without out-of-market carbon pricing payments to combat declining wholesale energy prices. \cite{HARATYK2017150} partially attributed premature retirements and the need for out-of-market support to high nuclear operating costs, defined as the sum of fuel, variable and fixed operating costs, which motivated this work to explore whether low SMR NPVs found by \cite{steigerwald2023} could be attributed to the same factors. Two key aspects of this work are distinct from \cite{HARATYK2017150}, who robustly modeled existing nuclear performance under historical wholesale prices, capacity markets, and policy mechanisms such as carbon pricing. First, this work models SMRs, which are purported to have lower investment costs from potential module-based learning \cite{inlatbinput}, and different operational characteristics such as ramping at a rate up to 1\% of rated thermal power per hour, as well as refueling modules sequentially \cite{alhadhrami2023}. Secondly, as wholesale power price trends are set to change with evolving grid complexity, such as increasing shares of Variable Renewable Energy (VRE) \cite{lbnlelecprices}, storage \cite{storage_elec_prices} and Electric Vehicles (EVs) \cite{SAUTER2024123002}, this work models future price scenarios to help capture these price trends in the analysis. 

Building on \cite{HARATYK2017150} and \cite{steigerwald2023}, the primary novel contribution this paper makes is identifying a new important mechanism not appreciated in previous literature --- that the economic performance and feasibility of SMR depends more on their marginal costs, defined as the sum of fuel and variable operating costs, rather than oft-discussed investment costs. Additional contributions this works makes are capturing the expanding set of revenue streams available under current market structures and policy incentives, and determining their necessary levels to make SMRs economically feasible. Fuel costs for SMRs are also calculated from manufacturer provided design parameters to add to the literature around SMR cost estimates. The final novel contribution is quantifying the breakeven marginal and investment cost tradeoff for SMRs under various scenarios. From this, investment cost ceilings are extracted as benchmarks to contribute to the discussion around the economics of SMRs.

Overall, the results of this work show that an environment of prices and subsidies beyond historic highs is necessary to attract private investment at manufacturer advertised cost benchmarks for SMRs. Primarily, the poor performance of SMRs can be attributed to \textbf{the consequences of high operating costs} in the transition to advanced nuclear. In an effort to reduce upfront investment cost for SMRs, design tradeoffs such as using novel High-Assay Low-Enriched Uranium (HALEU) and Tri-structural Isotropic (TRISO) fuels that do not have established supply chains ballooned operational and marginal costs. Intra-hour losses from operating the SMRs as a result overshadow any upfront investment cost advantages. While SMRs may potentially experience issues such as loss of economies of scale \cite{KUZNETSOV2008242} and other risks of new technology, this work demonstrates the greater importance of operating and marginal costs towards private investment feasibility.

High marginal costs at manufacturer advertised cost estimates imply that SMRs relying only on wholesale market revenues are unable to achieve discounted payback periods in less than their technical lifetimes. Improving feasibility with supplementary capacity revenues requires average lifetime prices to be greater than PJM's historic high 2025/26 auction price, which caused the Commonwealth of Pennsylvania to threaten to leave PJM \cite{Howland2025_PJMStatesGovernance}. Any potential policy support needs to exceed current levels of the Investment Tax Credits (ITC) and Production Tax Credits (PTC), and even with capacity revenues, a high PTC rate is needed, with capacity revenues more impactful than the PTC. For some SMRs, even if the ITC subsidized the whole cost of construction, it is still insufficient to improve payback periods below technical lifetimes. However, if policy makers were to raise the PTC to $\frac{\$95}{MWh}$, to better reflect the social cost of carbon \cite{socialcostofcarbon} of avoided emissions (primarily natural gas at roughly 0.5 tons/MWh), most SMRs would be able to achieve payback periods below 20 years, with or without capacity revenues. Quantifying the tradeoff between marginal and investment costs for SMRs, improvements in marginal costs pay larger dividends than investment cost reductions. If SMR marginal costs improved to baseload levels at $\frac{\$12}{MWh}$, investment cost benchmarks can be relaxed by an average of 30\%, with a maximum investment cost of $\frac{\$7,813}{kWe}$ under high wholesale and capacity prices supplemented by high PTC rates. Finally, if SMRs experience investment cost escalations, they have similar profitability to Vogtle 3 \& 4, with SMRs at current estimates as profitable as Nth-of-a-kind (NOAK) and advanced estimates of the AP1000. This sharply brings into question the benefits of the transition to advanced nuclear plants, especially since traditional plants enjoy economies of scale that SMRs lack.

Following this introduction section, Section \ref{sec::methods} describes the SMR cost and operational dataset, the price profiles used for energy market analysis and the dispatch algorithm. Section \ref{sec::results} presents the results in five parts. First, Section \ref{sec::energyresults} presents the results of energy markets only under future and historical price scenarios, Section \ref{sec::capacitymarket} then analyzes the impact of capacity revenues and Section \ref{sec::policysupport} presents policy support results. Section \ref{sec::tradeoffcurves} then presents results from breakeven marginal and investment cost tradeoff, and Section \ref{sec::sensitivities} presents results from benchmarking to traditional nuclear power and potential cost overruns.

\section{Data and Methodology}
\label{sec::methods}
\subsection{Data Collection}
\begin{figure*}[b!]
    \centering
    \includegraphics[width=\textwidth]{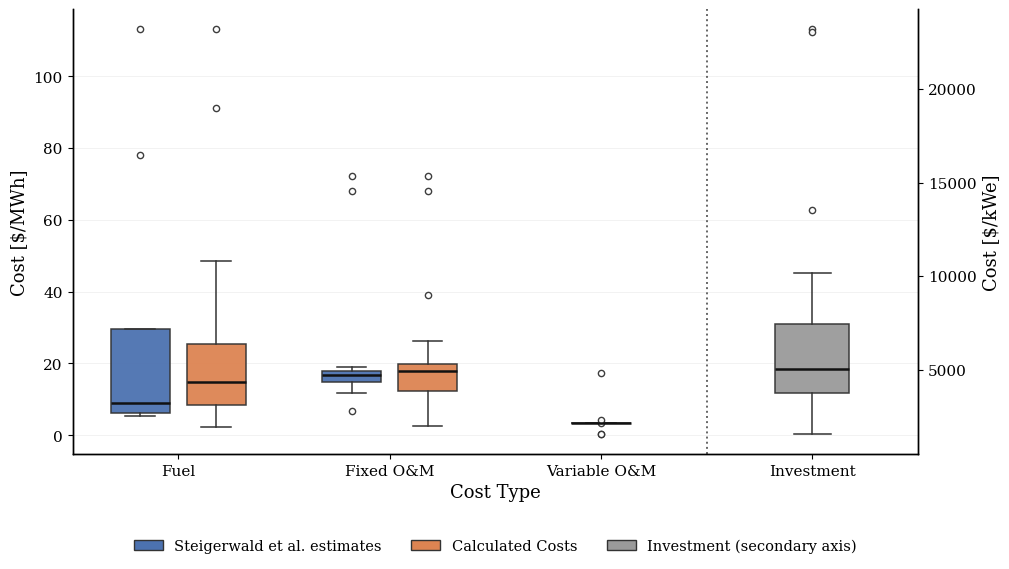}
    \caption{Cost estimates used by Steigerwald et al. and calculated in this work. Fixed costs were converted from $\$/\mathrm{MWe}$ to $\$/\mathrm{MWh}$ for ease of comparison.}
    \label{fig:cost-breakdown}
\end{figure*}

This paper utilizes a cost dataset from Steigerwald et al. \cite{steigerwald2023}. Investment costs were maintained, while operational data from the International Atomic Energy Agency's Advanced Reactors Information System (ARIS) \cite{IAEA_ARIS} were added, along with manufacturer-provided information for concepts not included in the original dataset. Although manufacturer-advertised investment costs may reflect optimistic assumptions and remain subject to uncertainty, this work does not assess their validity in detail, as the analysis primarily focuses on the role of marginal costs in SMR economic performance. Lazard assumptions for Variable O\&M data of $\frac{\$3.55}{MWh}$ \cite{lazard2024lcoe} were used for SMR concepts without otherwise available estimates of variable O\&M costs. For SMRs with missing fixed cost values, EIA parameters were used \cite{EIA2025CapitalCost}. Additionally, assumptions for SMR concepts using online refueling were taken as 6-12 months. Prices were normalized to 2020 USD for this work to maintain consistency. Figure \ref{fig:cost-breakdown} illustrates the differences between the dataset used in this work and Steigerwald et. al.

\begin{table}[t!]
\centering
\caption{Selected manufacturer-advertised cost data taken from the source data of \cite{steigerwald2023} and expanded using \cite{lazard2024lcoe}. Xe-100 estimates were obtained from \cite{Xenergy_AAC_2022} and Aurora estimates from \cite{oklo2023investor}.}
\label{tab:smrselectedcostdata}

\scriptsize
\setlength{\tabcolsep}{1.5pt}
\renewcommand{\arraystretch}{1.15}

\begin{tabular}{
  @{}
  >{\raggedright\arraybackslash}p{0.17\columnwidth}
  >{\raggedleft\arraybackslash}p{0.10\columnwidth}
  >{\raggedleft\arraybackslash}p{0.16\columnwidth}
  >{\raggedleft\arraybackslash}p{0.13\columnwidth}
  >{\raggedleft\arraybackslash}p{0.17\columnwidth}
  >{\raggedleft\arraybackslash}p{0.15\columnwidth}
  @{}
}
\toprule
\parbox[t]{0.17\columnwidth}{\raggedright Project} &
\parbox[t]{0.10\columnwidth}{\centering Capacity\\{[MWel]}} &
\parbox[t]{0.16\columnwidth}{\centering Investment\\Cost\\{[\$/MWel]}} &
\parbox[t]{0.13\columnwidth}{\centering Fuel\\Cost\\{[\$/MWh]}} &
\parbox[t]{0.17\columnwidth}{\centering Fixed O\&M\\Cost\\{[\$/MW-yr]}} &
\parbox[t]{0.15\columnwidth}{\centering Variable O\&M\\Cost\\{[\$/MWh]}} \\
\midrule
BWRX-300  & 300 & 2,250,000 & 8.44  & 140,160 & 3.55 \\
UK-SMR    & 470 & 5,215,937 & 10.95 & 595,680 & 3.55 \\
SMR-160   & 160 & 6,312,500 & 14.48 & 103,631 & 3.55 \\
NuScale   & 77  & 3,466,000 & 17.42 & 632,998 & 3.55 \\
Aurora-15 & 15  & 3,800,000 & 13.41 & 135,868 & 4.25 \\
Xe-100    & 80  & 7,500,000 & 3.39  & 21,900  & 3.55 \\
\bottomrule
\end{tabular}

\end{table}

\begin{table}[t!]
\centering
\caption{Operational data for SMR concepts from IAEA's ARIS \cite{IAEA_ARIS}. Xe-100 estimates were obtained from \cite{Xenergy_AAC_2022} and Aurora estimates from \cite{oklo2023investor}.}
\label{tab:smrselectedopdata}

\scriptsize
\setlength{\tabcolsep}{3pt}
\renewcommand{\arraystretch}{1.1}

\begin{adjustbox}{max width=\columnwidth,center}
\begin{tabular}{lrrrrr}
\toprule
Project & Lifetime [yr] & Modules & Const. [mo.] & Refuel Min. [mo.] & Refuel Max. [mo.] \\
\midrule
BWRX-300 & 60 & 1  & 26 & 12  & 24  \\
UK-SMR   & 60 & 1  & 24 & 18  & 24  \\
SMR-160  & 80 & 2  & 36 & 20  & 26  \\
NuScale  & 60 & 12 & 36 & 15  & 18  \\
Aurora-15 & 40 & 1 & 12 & 118 & 120 \\
Xe-100 & 60 & 4 & 36 & 6 & 12 \\
\bottomrule
\end{tabular}
\end{adjustbox}

\end{table}

Front-end fuel costs were calculated using the Argonne National Laboratory NE-COST / CNPCE calculator \cite{ANL_CNPCE_QNECost} after extracting manufacturer advertised design parameters from ARIS \cite{IAEA_ARIS}. Fabrication cost estimates from Rothwell \cite{ROTHWELL2010538} were used after cross referencing publicly available literature on fuel supply sources for each concept. HALEU and advanced fuel enrichment and deconversion costs were taken from White and Cothran \cite{white2023haleu} and enrichment, fabrication, depleted uranium deconversion, spent nuclear fuel conditioning, and geologic disposal costs were taken from the DOE Fuel Cycle Evaluation and Screening \cite{doe2014fuelcycleappendixc}. OECD-NEA's back-end fuel-cycle cost studies \cite{nea2013backend} were used for sensitivities. A 92.3\% capacity factor was used for fuel cost calculations \cite{Mueller2025GenerationCapacity}. 

For the purposes of this work, we focus on six manufacturer concepts - GE-Hitachi's BWRX-300, Rolls-Royce's UK-SMR \footnote{Although its 470 MWe capacity exceeds the traditional International Atomic Energy Agency (IAEA) limit of 300 MWe, the Rolls-Royce design is classified as a Small Modular Reactor (SMR) due to its reliance on standardized factory prefabrication and modular construction methodologies.}, Holtec's SMR-160, NuScale's VOYGR, OKLO's Aurora 15MWe Powerhouse, and X-energy's Xe-100. These are widely considered as some of the leading concepts \cite{SCHLEGEL2024283} and performed the best in our analysis. Table \ref{tab:smrselectedcostdata} outlines the cost data selected for the bulk of this paper's analysis, with Table \ref{tab:smrselectedopdata} outlining the respective operational data. While fixed costs were converted to \$/MWh to compare against operating costs in Figure \ref{fig:cost-breakdown}, assuming  they were not included in dispatch decisions, which are only based on marginal costs.

\subsection{Price Profiles}

\begin{figure*}[b!]
    \centering
    \includegraphics[width=0.8\textwidth]{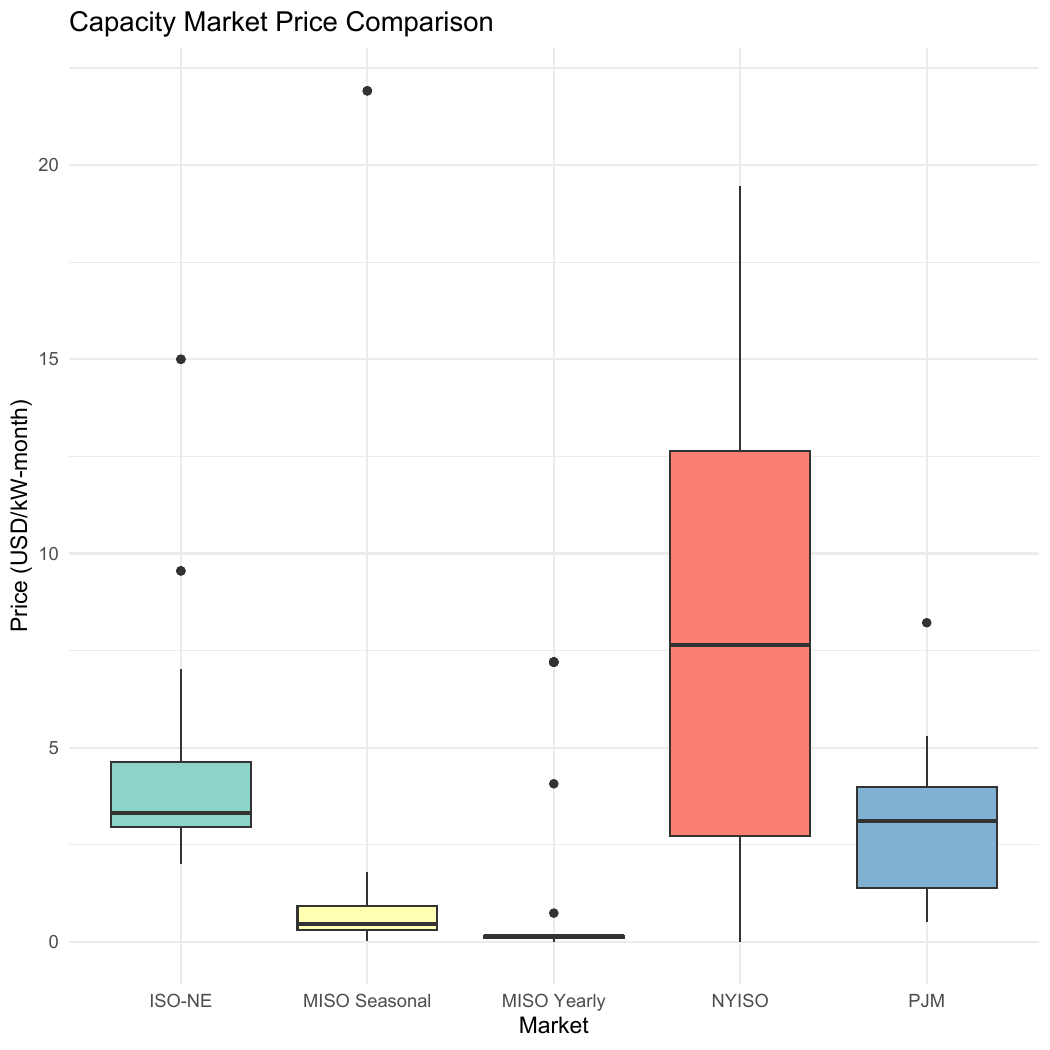}
    \caption{Comparison of historical prices of existing capacity markets in MISO, ISO-NE, NYISO, and PJM. Prices include the PJM 2025/26 capacity market auction, shown by the outlier at \$8.21/kW-month.}
    \label{fig:capacity-prices-breakdown}
\end{figure*}

Ten hourly real-time price profile scenarios from the National Renewable Energy Laboratory's Cambium 2022 and 2023 dataset \cite{cambium2023documentation}, spanning 2025-2050, were used to represent future prices. To benchmark against historic scenarios, hourly prices between 2011-2023 from six competitive wholesale markets were extracted from Lawrence Berkeley National Laboratory's Renewables and Wholesale Electricity Prices dataset (ReWEP) \cite{lbnlhistoricalprices}. Element-wise interpolation was then used to link Cambium profiles five years apart, to create a single profile scenario up to the SMR technical lifetime. While all SMR concepts were run in all scenarios, this work presents results from three future and three historic price scenarios, detailed in Table \ref{tab:scenariosdata}. Additionally, capacity market revenues were modeled as monthly payments at a fixed rate and discounted over time. Capacity prices were varied at and beyond historical limits shown in Figure \ref{fig:capacity-prices-breakdown}. 

\begin{table}[H]
\centering
\caption{Summary statistics of highlighted historical and future price scenarios used in the study. Three Cambium scenarios and three average ISO historical price scenarios were selected to contrast performance expectations. All prices are in $\$/\mathrm{MWh}$.}
\label{tab:scenariosdata}

\scriptsize
\setlength{\tabcolsep}{3pt}

\resizebox{\columnwidth}{!}{%
\begin{tabular}{llrrrr}
\toprule
Price Scenario & Type & Mean & Std. Dev. & Max & Min \\
\midrule
High Natural Gas Prices      & Future   & 38.53 & 101.66 & 2455.31 & -192.27 \\
Mid Case 100                 & Future   & 55.24 & 130.35 & 3221.74 & -733.07 \\
High Renewable Energy Cost   & Future   & 36.87 & 103.58 & 2338.17 & -148.10 \\
PJM                          & Historic & 41.51 & 48.30  & 4002.80 & -284.60 \\
ERCOT                        & Historic & 41.77 & 292.85 & 9080.56 & -144.54 \\
CAISO                        & Historic & 44.56 & 48.91  & 1600.10 & -2491.30 \\
\bottomrule
\end{tabular}%
}
\end{table}

\begin{table}[H]
\centering
\caption{Future cost estimates of the AP1000 and NREL ATB reactor scenarios.}
\label{tab:research_estimates}

\fontsize{7.5}{9}\selectfont
\setlength{\tabcolsep}{2pt}
\renewcommand{\arraystretch}{1.15}

\begin{tabular}{
  @{}
  >{\raggedright\arraybackslash}p{0.27\columnwidth}
  >{\raggedleft\arraybackslash}p{0.17\columnwidth}
  >{\raggedleft\arraybackslash}p{0.13\columnwidth}
  >{\raggedleft\arraybackslash}p{0.17\columnwidth}
  >{\raggedleft\arraybackslash}p{0.15\columnwidth}
  @{}
}
\toprule
\parbox[t]{0.27\columnwidth}{\raggedright Concept} &
\parbox[t]{0.17\columnwidth}{\centering Construction\\Cost\\{[\$/kWel]}} &
\parbox[t]{0.13\columnwidth}{\centering Fuel\\Cost\\{[\$/MWh]}} &
\parbox[t]{0.17\columnwidth}{\centering Fixed O\&M\\Cost\\{[\$/MW-yr]}} &
\parbox[t]{0.15\columnwidth}{\centering Variable O\&M\\Cost\\{[\$/MWh]}} \\
\midrule
Baseline (V3\&4 realized)        & 11,000 & 9.00  & 204,000 & 3.40 \\
Baseline (V3\&4 if built today)  & 15,000 & 9.00  & 204,000 & 3.40 \\
Next 2 @ Vogtle                  & 8,300  & 9.00  & 175,000 & 2.80 \\
Next 2 @ Greenfield              & 9,300  & 9.00  & 175,000 & 2.80 \\
NOAK                             & 4,625  & 9.00  & 126,000 & 1.90 \\
ATB\_LR\_Adv                     & 6,471  & 9.13  & 126,000 & 1.90 \\
ATB\_LR\_Mod                     & 7,616  & 10.18 & 17,000  & 2.80 \\
ATB\_LR\_Cons                    & 11,836 & 11.34 & 204,000 & 3.40 \\
\bottomrule
\end{tabular}

\end{table}

\subsection{Policy and Research Estimates}

Investment and production tax credits were modeled to calculate policy impacts  on nuclear generators. Investment credits were treated as cost reduction multipliers on the investment costs, and production as an hourly credit spanning 10 years, based on a reference of the ITC and PTC. PTC and ITC rates were taken from NREL's Annual Technology Baseline (ATB) \cite{inlatbinput}, with ITC cost reduction multipliers  calculated from investment cost reductions in the ATB. Additionally, research estimates from the ATB and the 2024 Total Cost Projection of Next AP1000 \cite{ap10002024} were used to benchmark industry expectations on cost data for SMRs and AP1000's. Table \ref{tab::research_estimates} details the collected estimates.

\subsection{Flexible Operations Model}

Using compiled price profiles and generator data, a simple Flexible Power Operation (FPO) price taker formulation was used to simulate reactor performance over its lifetime. The relation of the hourly electricity price to the marginal cost of operation was the central objective of the FPO. If the cost was lower than the hourly price, the SMR was expected to be operate at maximum power. If the cost was higher, the SMR was expected to ramp down to a technical Low Power Output (LPO) defined at 40\% of maximum power \cite{alhadhrami2023}.

Refueling was scheduled one module at a time, occurring when the price signal is in the lower quartile of the refueling range for the generator. Startup costs were obtained from the ATB at $\frac{\$60}{kWe}$ \cite{inlatbinput}. Capacity payments were treated as monthly injections exogenous of the dispatch, but included in discounting. From this, the price taker formulation was defined as follows, starting with \(P(t)\) as the net operating cashflow.

\[
\resizebox{\columnwidth}{!}{$\displaystyle
P(t)=
\begin{cases}
  p(t)m + I_C(t)cm - (f+v)m, 
  & \text{if } p(t)\ge f+v,\ s(t)=1,\ t\in[t_s,t_e],\\[4pt]
  p(t)m' + I_C(t)cm' - (f+v)m' - SC, 
  & \text{if } p(t)\ge f+v,\ s(t)=0,\ t\in[t_s,t_e],\\[4pt]
  p(t)m - (f+v)m, 
  & \text{if } p(t)\ge f+v,\ s(t)=1,\ t\notin[t_s,t_e],\\[4pt]
  p(t)m' - (f+v)m' - SC, 
  & \text{if } p(t)\ge f+v,\ s(t)=0,\ t\notin[t_s,t_e],\\[4pt]
  p(t)\ell + I_C(t)c\ell - (f+v)\ell, 
  & \text{if } p(t)< f+v,\ s(t)=1,\ t\in[t_s,t_e],\\[4pt]
  p(t)\ell + I_C(t)c\ell - (f+v)\ell - SC, 
  & \text{if } p(t)< f+v,\ s(t)=0,\ t\in[t_s,t_e],\\[4pt]
  p(t)\ell - (f+v)\ell, 
  & \text{if } p(t)< f+v,\ s(t)=1,\ t\notin[t_s,t_e],\\[4pt]
  p(t)\ell - (f+v)\ell - SC, 
  & \text{if } p(t)< f+v,\ s(t)=0,\ t\notin[t_s,t_e].
\end{cases}
$}
\]

Where \(m\) total plant capacity, \(m'\) is the refueling plant capacity, \(\ell\) is the LPO, \(p(t)\) denotes the hourly electricity price at time \(t\),\(f\) and \(v\) correspond to the fuel cost and variable operating cost, respectively, \(c\) represents the hourly production tax credit and startup cost is denoted by \(SC\). The binary functions \(s(t)\in\{0,1\}\) indicate the refueling status at time \(t\) and \(I_C(t)\) denotes the production credit eligibility, between the operational start time \(t_s\) and credit eligibility expiration \(t_e\).

Key SMR performance metrics of NPV and Discounted Payback Period were collected. Investment costs were distributed across the construction duration with a 10\% construction interest applied \cite{PORTUGALPEREIRA2018158}. For the purposes of this work, a 4\% discount rate was assumed in all Present Value calculations. Using these assumptions, the NPV and Discounted Payback Period was formulated. Let the annual outlay \(I_t\) be:

\[
\quad
I_{t} \;=\; \frac{C_{\mathrm{const}}}{D}\,\bigl(1 + r_{c}\bigr)^{\,t-1},
\quad t = 1,\dots,D.
\]

where \(C_{\mathrm{const}}\) is the total construction cost, \(D\) is the total construction duration in years, and \(r_c\) is the construction interest rate applied during this period.

The present value of investment \(I_0\) is then defined as:

\[
\quad
I_{0}
\;=\;
\sum_{t=1}^{D}
\frac{I_{t}}{(1 + r)^{t}}
\;=\;
\sum_{t=1}^{D}
\frac{\displaystyle\frac{C_{\mathrm{const}}}{D}\,\bigl(1 + r_{c}\bigr)^{\,t-1}}
{\,(1 + r)^{t}}.
\]

where \(r\) is the discount rate used for present-value calculations.

From this, the net present value is defined as:
\[
\mathrm{NPV}
\;=\;
-\,I_{0}
\;+\;
\sum_{t=1}^{N}
\frac{P_{t} - FOM}{(1 + r)^{t}}.
\]

where \(P_t\) is the net operating cashflow in year \(t\), \(FOM\) is the Fixed O\&M cost and \(N\) is the technical lifetime of the plant.

Finally, the discounted payback period \(T\) is defined as the smallest integer such that the cumulative discounted operating cash flows satisfy \(\sum_{t=1}^{T} P_t/(1+r)^t \ge I_0\), i.e.:

\[
\min \Biggl\{\,T\in\{1,\dots,N\}\;\Bigm|\;
\sum_{t=1}^{T}
\frac{P_{t}}{(1 + r)^{t}}
\;\ge\;
I_{0}
\Biggr\}.
\]

\section{Results}
\label{sec::results}
\subsection{Investment Viability in Competitive Energy Markets}
\label{sec::energyresults}
Exploring the private investment viability of Small Modular Reactors, this paper first finds that energy only market revenue is insufficient under nearly all historical and future price scenarios. Almost all concepts have discounted payback periods greater than their technical lifetimes. The only reactor to show promise in several of the scenarios, PBMR-400, was canceled as a demonstration project by the South African government due to a lack of financing opportunities, being considered uneconomical and having accumulated significant losses \cite{pbmrcancel}. Observing the scenarios themselves, SMRs necessitate an environment of high decarbonization and high prices to achieve feasible economic outcomes. This is seen by the performance of SMRs in the Mid Case 100 scenario, a scenario with full decarbonization, mean prices nearly $\frac{\$15}{MWh}$ higher than the next closest scenario and significantly more price volatility, as observed in Table \ref{tab:scenariosdata}.

\begin{figure*}[b!]
  \centering
  \begin{tabular}{@{}ccc@{}}
    \includegraphics[width=0.32\textwidth]{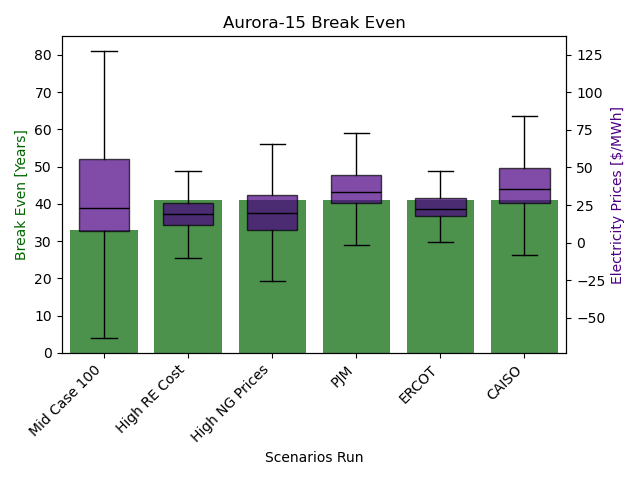} &
    \includegraphics[width=0.32\textwidth]{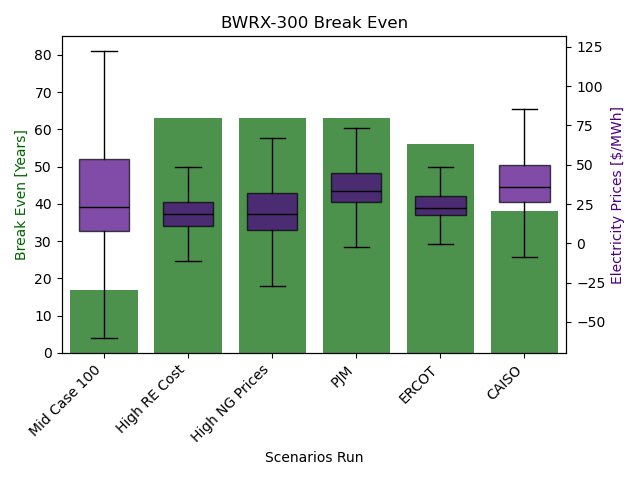} &
    \includegraphics[width=0.32\textwidth]{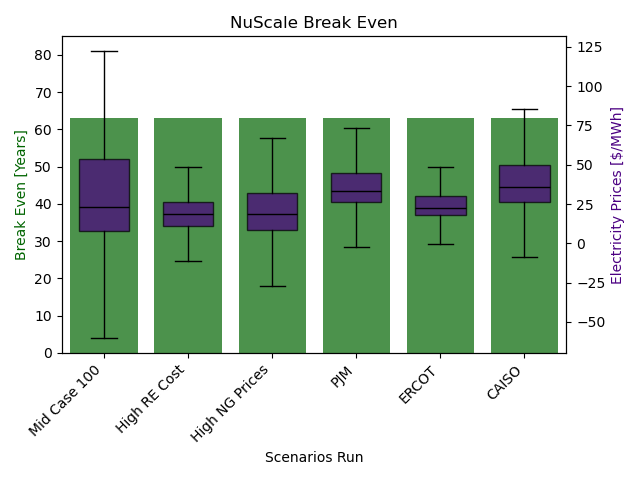} \\[1ex]
    \includegraphics[width=0.32\textwidth]{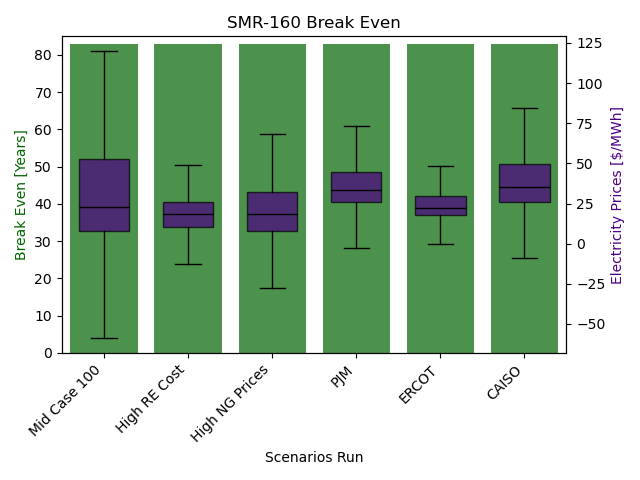} &
    \includegraphics[width=0.32\textwidth]{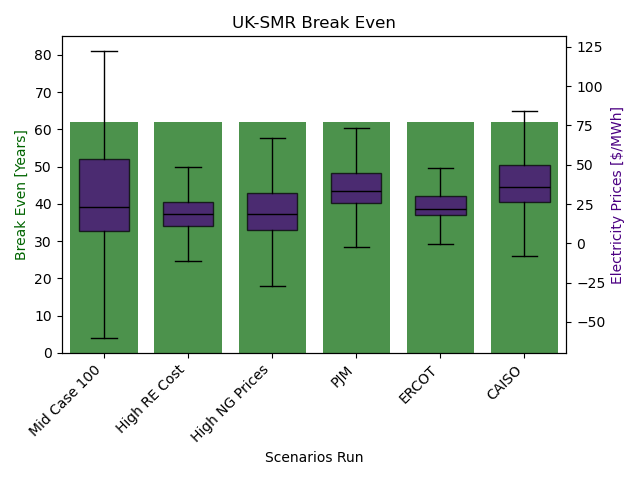} &
    \includegraphics[width=0.32\textwidth]{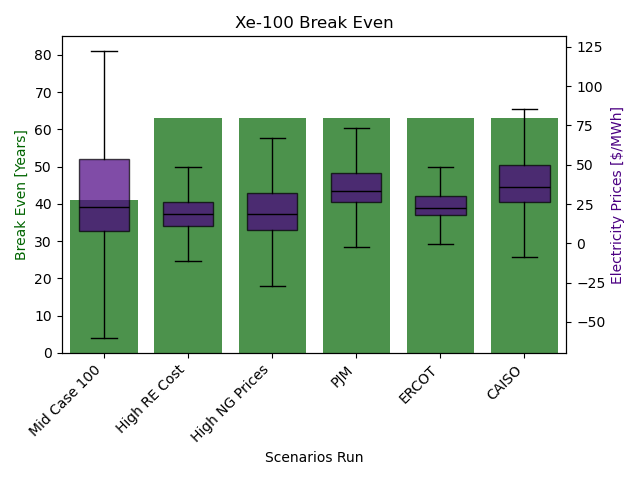}
  \end{tabular}
  \caption{Discounted payback period (green) of six market-leading and top-performing reactors, with the distribution of electricity prices (purple) overlaid for selected scenarios.}
  \label{fig:2x3grid}
\end{figure*}

The key finding of this paper is that high marginal and operating costs drive the poor performance of SMR concepts. Wholesale energy markets set the system energy price by the merit order based on short-run marginal cost. As such, due to ramping and startup constraints forcing operation, SMRs absorb significant intra-hour losses during hours where price depression occurs and their marginal costs are higher than the system price. These losses are so large that even if investment cost reductions are achieved, SMRs would still be unprofitable. An example is the case of the Rolls Royce's UK-SMR. With reported fixed operating costs of $\frac{\$68}{MWh}$ \cite{PowerTech2025UKFirstSMR} and marginal costs of $\frac{\$14.5}{MWh}$, the concept endures significant intra-hour losses which, over the course of its operation, results in a discounted payback period greater than its technical lifetime, even if it were to incur no investment costs. Additionally, during the hours where the market clearing prices are higher than the marginal costs, SMRs earn smaller inframarginal rents. As a result, even if SMRs are able to avoid technical constraints and instantaneously ramp from zero to maximum power to avoid intra-hour losses, compressed margins due to high marginal costs would still inhibit the profitability of a SMR plant. This is demonstrated in our results, where the most substantial reduction to discounted payback periods from bypassing technical constraints is only 2 years, achieved by the BWRX-300 under the Mid-Case 100 price scenario.  

These dynamics are reflected in operational decisions made by SMRs. SMRs are, like traditional nuclear power, expected to provide baseload power to the power grid \cite{DOE_AdvancedNuclear_2025}. However, as seen in Table \ref{smr_capacity_factors}, in order to improve economic prospects by minimizing intra-hour losses, SMRs are more likely to utilize flexible ramping features than operate as baseload power, dropping average lifetime capacity factors to 47.4\% in some price scenarios. In the price scenario Mid Case 100 where SMRs perform the best, capacity factors average 78.7\% across reactors, significantly lower than their expected values of 92\% \cite{doesmrcapacityfactors}, and of current baseload nuclear power plants at 93.4\% \cite{DOEcapacityfactors}. Additionally, if SMRs were able to ramp to zero power, capacity factors would drop further by 7\% in most scenarios. At this level, SMRs would be operating at similar capacity factors as rapid ramping Natural Gas plants, which have average capacity factors of 59.9\% \cite{DOEcapacityfactors}. 

\begin{table*}[t!]
  \centering
  \caption{Capacity factors for different SMR designs in selected scenarios from Cambium and ISO historical prices. Values are shown as percentages.}
  \label{tab:smr_capacity_factors}

  \small
  \setlength{\tabcolsep}{6pt}
  \renewcommand{\arraystretch}{1.12}

  \begin{tabular*}{\textwidth}{@{\extracolsep{\fill}}lrrrrrr@{}}
    \toprule
    & \multicolumn{6}{c}{Capacity factor [\%]} \\
    \cmidrule(l){2-7}
    Scenario & BWRX-300 & UK-SMR & SMR-160 & NuScale & Aurora-15 & Xe-100 \\
    \midrule
    Mid Case 100   & 62.1 & 81.8 & 81.7 & 80.7 & 81.2 & 84.7 \\
    High RE Cost   & 47.4 & 85.3 & 84.7 & 83.6 & 85.2 & 89.5 \\
    High NG Prices & 52.9 & 80.8 & 80.1 & 79.3 & 80.9 & 84.9 \\
    PJM            & 68.8 & 99.6 & 99.6 & 99.5 & 99.5 & 99.8 \\
    ERCOT          & 51.3 & 97.2 & 97.2 & 96.7 & 96.9 & 98.3 \\
    CAISO          & 72.4 & 96.3 & 96.3 & 96.2 & 96.2 & 96.9 \\
    \bottomrule
  \end{tabular*}
\end{table*}
Evidence suggests that the high SMR marginal costs discussed so far may be a trade-off by manufacturers in an effort to reduce upfront investment costs. A majority of concepts experience higher fuel costs than traditional reactors as they use higher enrichment fuels such as HALEU, or need significant investments into new supply chains such as TRISO fuels \cite{LARSEN2025101876, IAEA2023SMRFuelSupply, Reuters2025SMRFuel}. This is observed in this work through average SMR fuel costs in the compiled dataset of manufacturer provided costs being $\frac{\$23.69}{MWh}$, significantly higher than traditional nuclear reactors utilized as baseload power experiencing costs between $\frac{\$9}{MWh}$ and $\frac{\$11}{MWh}$. Literature also shows that concepts using traditional fuels such as NuScale are expected to have fuel costs 15\% to 70\% higher than traditional reactors \cite{pannier2014}, pointing to a possible industry-wide trend. Additionally, with higher fixed costs averaging $\frac{\$21.81}{MWh}$, observable in Figure \ref{fig:cost-breakdown}, operating and fuel costs of SMRs total to $\frac{\$45.5}{MWh}$, significantly higher than the average prices of all historical and non-decarbonization based scenarios except for Mid Case 100 in Table \ref{tab:scenariosdata}. A possible explanation for high SMR operating costs is that these concepts were designed for a high price environment, with NuScale's early designs expecting wholesale energy prices of $\frac{\$75}{MWh}$ \cite{ingersoll2014nuscale}, and OKLO's Aurora-15 expecting prices of $\frac{\$105}{MWh}$ \cite{oklo2023investor}.

\subsection{Investment Viability with Energy and Capacity Revenues}
\label{sec::capacitymarket}

\begin{figure*}[ht]
    \centering
    \includegraphics[width=\linewidth]{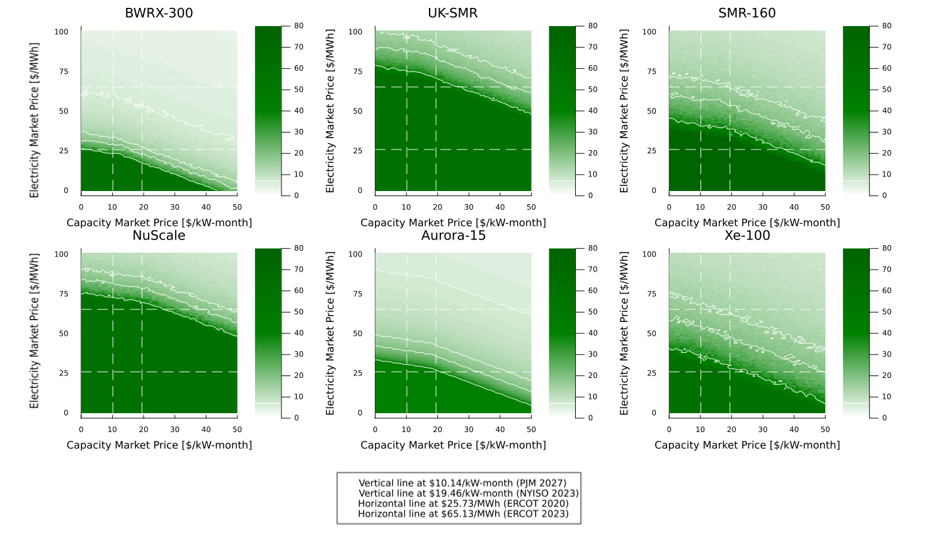}
    \caption{Discounted Payback Periods, dark green corresponding to 80 years, of selected SMR prototypes as a function of average energy and capacity prices.}
    \label{fig:energy-cap-heatmap}
\end{figure*}

Evaluating economic performance solely on dispatching to energy markets runs into the ``missing money problem" - where the prices for energy do not reflect the true investment and operational costs necessary for reliable power grid operation \cite{HOGAN201755}. While dispatchable resources can avail themselves of ancillary services revenue in some ISOs/RTOs, previous work has shown that co-bidding strategies with energy and ancillary service markets result in a 4\% increase in yearly revenues under historical CAISO prices \cite{alhadhrami2023}. This translates to an average impact of 0.43 years on discounted payback periods. Capacity markets, on the other hand, are a construct used to mitigate the missing money problem \cite{PNNL2022CapacityMarkets}, and provide a rich source of potential additional revenue for thermal generators. Recent rises in capacity prices \cite{Howland2025PJMCapacityPrices, Walton2024NECapacityPrices, Howland2025MISOCapacityPrices}, such as PJM's 2025/26 capacity prices soaring to $\frac{\$269.92}{MW-day}$ from $\frac{\$28.92}{MW-day}$ \cite{Howland2024PJMCapacityAuction}, translating to $\frac{\$8.21}{kW-month}$ and $\frac{\$0.88}{kW-month}$ respectively, means that a key factor in the private investment viability of Small Modular Reactors are the role of capacity revenues. 

As such, this work explores how current capacity payment levels would impact private investment feasibility, and what average lifetime price levels would be necessary to overcome intra-hour energy market losses from high operating costs. To simulate this, average lifetime electricity prices that do not contain temporal variability and volatility features of the electricity price profiles used in Section \ref{sec::energyresults} were used.

Results show that the dominant sensitivity of SMR payback periods is to wholesale energy prices. If capacity prices range around historical levels, as illustrated in Figure \ref{fig:capacity-prices-breakdown}, revenues are insufficient to overcome intra-hour energy market losses, and offer minimal improvements to payback periods. As such, this work finds that for capacity payments to offer noticeable improvements in payback periods, average lifetime prices need to significantly exceed historic highs hit in PJM's 2027/28 auction of $\frac{\$10.14}{kW-month}$.

Determining what combination of electricity price and capacity price is necessary for SMRs to find private investment feasibility is highly contextual. SMR concepts vary in cost structures, construction periods, refueling time and number of modules. As such, capacity and electricity price benchmarks should be evaluated on the basis of the SMR concept. Analyzing the performance of current SMR concepts, most are still unable to achieve payback periods of 20 years or less within the defined quadrant of economic feasibility based on historical price data. 12 out of the 22 concepts do not achieve payback periods of less than 20 years even if average lifetime prices were at ERCOT's 2023 price of $\frac{\$65.13}{MWh}$ and NYISO's 2023 capacity price of $\frac{\$19.46}{kW-month}$. Intra-hour energy market losses from high operating costs are so high for the CEFR and CAREM SMR concepts that they are unable to achieve payback periods below their technical lifetimes even if capacity and energy prices reached $\frac{\$100}{kW-month}$ and $\frac{\$100}{MWh}$, respectively. 

Considering that the average payback period of a solar PV or onshore wind plant is between 2-20 years \cite{windsolarpaybackperiod}, SMRs are less likely to be competitive as an investment opportunity, even considering both energy and capacity revenues. This is punctuated further by the fact that SMRs avail themselves of greater proportions of capacity revenues due to their higher ELCC's (Effective Load Carrying Capacity), essentially the assumed capacity factor, of around 95\% compared to 10-60\% for wind and solar \cite{johnson2024elcc}. It should be noted that 2-20 year payback periods for wind and solar systems are including tax credits, so whether SMRs would be able to achieve similar economic performance with the inclusion of tax credits is the focus of the following section. Particularly of interest is whether a combination of policy support along with energy and capacity revenues is sufficient to achieve private investment feasibility.

\subsection{Impacts of Policy Support}
\label{sec::policysupport}
Tax credits have been a centerpiece of financing nuclear power projects to improve private investment feasibility. In fact, most nuclear projects only become bankable for private investors after de-risking through government involvement \cite{WEIBEZAHN2024114382}. With the introduction of the Clean Electricity Investment Credit and Zero-Emission Nuclear Power Production Credit in the 2022 Inflation Reduction Act \cite{irs2025clean, irs2025nuclear}, SMRs in service after 2024 can choose to subsidize either a portion of their investment cost or receive an operational credit for up to 10 years.

\begin{figure*} [H]
    \centering
    \includegraphics[width=\linewidth]{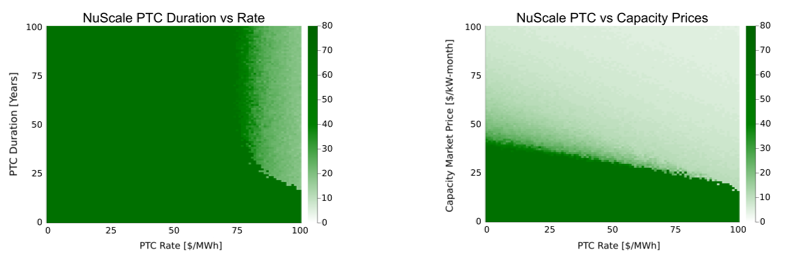}
    \caption{[Left] Discounted Payback Periods as a function of the duration and rate of the PTC. [Right] Discounted Payback Periods as a function of the PTC and Capacity Price.}
    \label{fig:ptc-payback}
\end{figure*}

Analyzing these policy support mechanisms, this work finds that maximum available levels of the investment tax credits, currently the 50\% rate, are insufficient to improve payback periods of most SMR concepts below their technical lifetimes in almost all scenarios. Additionally, the ITC would need to cover more than 75\% of investment costs for sub-20 year payback periods, and for some SMRs, even if tax credits covered the whole cost of construction, it is still insufficient to improve payback periods below technical lifetimes. This finding punctuates the impact of escalated operating costs on economic feasibility, where even zero investment costs of some SMRs such as Rolls Royce's UK-SMR and NuScale's VOYGR are unable to compensate for intra-hour losses.

Production tax credits (PTC) on the other hand, directly impact marginal cost competitiveness. Figure \ref{fig:ptc-payback} illustrates discounted payback periods as a function of PTC payout, eligibility period and capacity price. Private investment feasibility is defined as achieving sub-20 year discounted payback periods with PTC rates and capacity revenues at optimistic benchmarks. This is defined as at or under the maximum level of the PTC at $\frac{\$33}{MWh}$ for a period of 10 years, and average lifetime capacity prices under the historical high of the PJM 2027/28 auction price of $\frac{\$10.14}{kW-month}$. Mid Case 100, the optimistic price scenario in which SMRs performed best, was used to analyze PTC and capacity price impacts.

Observing the effects of the PTC only in Figure \ref{fig:ptc-payback}, two things are apparent. First, PTC duration beyond 10 years has nominal effects on payback periods, as most of the value is captured in the early years. Second, a threshold effect is observable with the rate of the PTC, where payback periods only improve after a certain PTC rate is achieved, with the rate depending on the SMR cost structure. Benchmarking to the feasibility standard defined earlier, this work finds that current levels of the PTC, if SMRs were able to claim them, are insufficient to make the majority of concepts attractive to private investment. While concepts such as GE-Hitachi's BWRX-300 and OKLO's Aurora-15 are able to achieve 20-year payback periods at or under Inflation Reduction Act (IRA) PTC levels, 18 out of 22 considered SMRs would not.

With the ability to capture capacity revenues in addition to production tax credits, average lifetime capacity prices have a larger influence than the PTC on feasibility. Duration of the PTC in particular becomes less influential with the availability of capacity revenues. However most SMRs, with the exception of a few reactors such as BWRX-300 and Aurora-15, necessitate capacity prices and PTC rates in excess of historic highs to be feasible for private investment. What should be noted is that if policy makers were to raise the PTC to $\frac{\$95}{MWh}$, to better reflect the social cost of carbon \cite{socialcostofcarbon} of avoided emissions (primarily natural gas at roughly 0.5 tons/MWh), most SMRs would be able to achieve payback periods below 20 years, with or without capacity revenues. For context, wind plants are found to be competitive at $\frac{\$21}{MWh}$ \cite{LU20114207}, with current base rates of the PTC starting at $\frac{\$30}{MWh}$ \cite{IRS2026CleanElectricityProductionCredit}.

\begin{figure*}[t!]
\centering

\includegraphics[width=\textwidth]{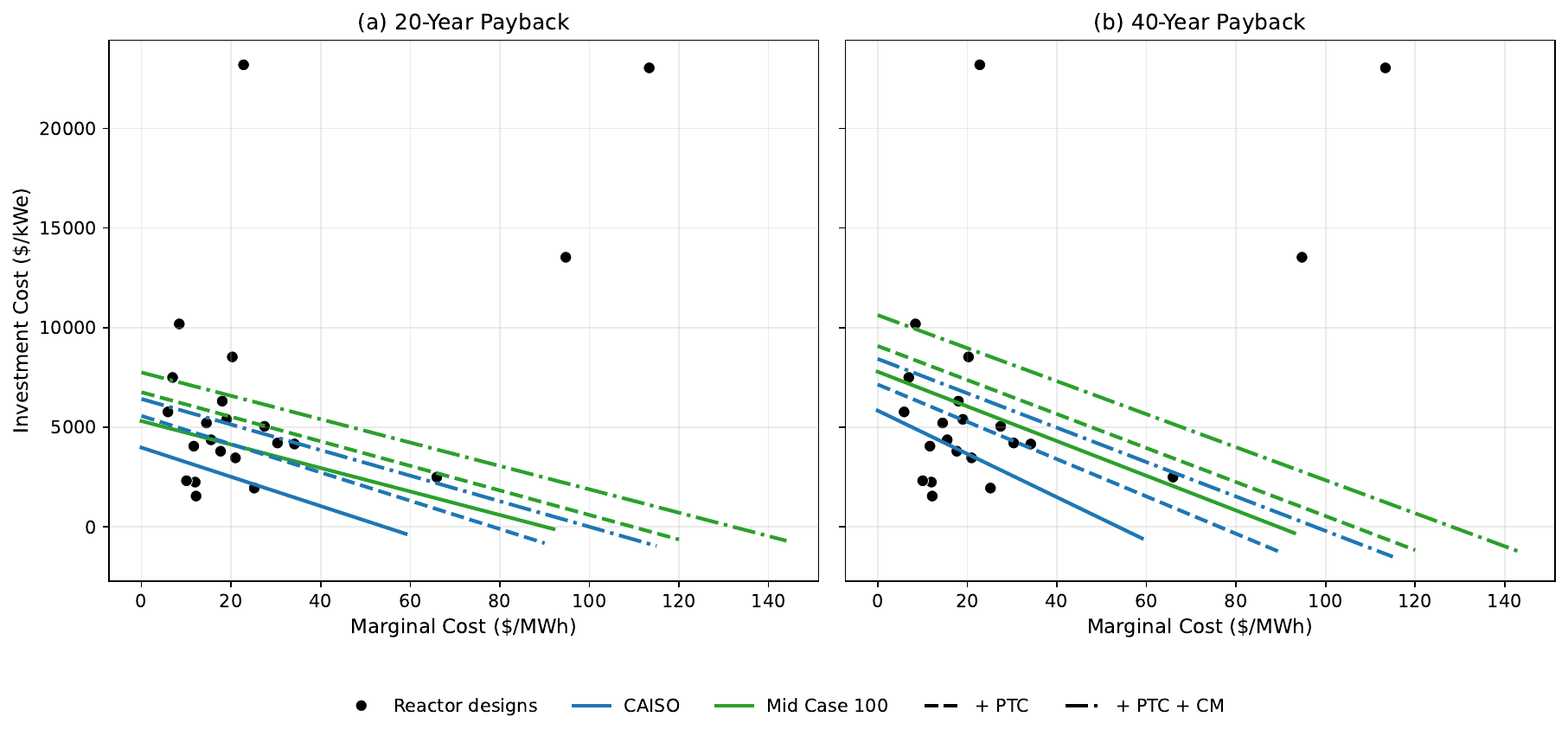}
\caption{Investment and marginal cost breakeven tradeoff curves for 20- and 40-year payback periods with current manufacturer costs as data points. Tradeoff curves are calculated for policy support (+PTC) and capacity revenues (+PTC + CM).}
\label{fig:figures_of_merit}

\vspace{1em}

\captionof{table}{Investment cost upper bounds to achieve a 20-year payback period under baseload marginal costs (\$12/MWh) and SMR manufacturer-advertised marginal costs (\$25/MWh).}
\label{tab:baseload_v_current_costsets}

\small
\setlength{\tabcolsep}{5pt}
\renewcommand{\arraystretch}{1.15}

\begin{tabular}{@{}p{0.58\textwidth}p{0.17\textwidth}p{0.17\textwidth}@{}}
\toprule
\textbf{Scenario} &
\centering\textbf{Baseload [\$/kWe]} &
\centering\arraybackslash\textbf{SMR [\$/kWe]} \\
\midrule
Mid Case 100 & \centering 4,699.71 & \centering\arraybackslash 3,936.77 \\
Mid Case 100 + PTC \$33/MWh & \centering 6,378.18 & \centering\arraybackslash 5,468.75 \\
Mid Case 100 + ITC 30\% & \centering 5,996.98 & \centering\arraybackslash 4,979.36 \\
Mid Case 100 + Capacity Rate \$10.14/kW-month & \centering 5,859.42 & \centering\arraybackslash 5,078.13 \\
Mid Case 100 + PTC + Capacity Rate & \centering 7,812.50 & \centering\arraybackslash 6,274.43 \\
\addlinespace[0.3em]
CAISO & \centering 3,222.66 & \centering\arraybackslash 2,050.78 \\
CAISO + PTC \$33/MWh & \centering 5,151.94 & \centering\arraybackslash 3,906.25 \\
CAISO + ITC 30\% & \centering 4,049.40 & \centering\arraybackslash 2,610.33 \\
CAISO + Capacity Rate \$10.14/kW-month & \centering 4,345.70 & \centering\arraybackslash 3,131.68 \\
CAISO + PTC + Capacity Rate & \centering 6,250.10 & \centering\arraybackslash 4,739.38 \\
\bottomrule
\end{tabular}

\vspace{0.4em}

\parbox{0.96\textwidth}{\scriptsize\textit{Reference costs:} Xe-100 (\$7,500/kWe), BWRX-300 (\$2,250/kWe), NuScale (\$3,466/kWe), Vogtle 3\&4 if built today (\$15,000/kWe), NOAK AP1000 (\$4,625/kWe), and Moderate AP1000 (\$7,616/kWe).}

\end{figure*}

For many concepts, SMR fixed costs are a significant determinant towards private investment feasibility. For example, NuScale's VOYGR advertises an annual fixed costs of \$185,000,000 for an 8-module plant \cite{ingersoll2014nuscale}, which alone overshadows projected energy market and supplementary capacity market revenues. This is also observed in the larger influence of capacity revenues as compared to production tax credits on economic feasibility. Capacity revenues supplement both investment and operating costs. As previous findings of this work showed investment subsidies having low impacts on payback periods, this suggests that the primary function of capacity revenues for most SMRs is reducing the burden of fixed costs. However, between fixed operating and marginal costs, most SMRs are still more sensitive to marginal costs due to their influence on bidding behavior into wholesale energy markets, which remain the largest revenue stream.

\subsection{Breakeven investment and marginal costs tradeoffs}
\label{sec::tradeoffcurves}

This work has thus far explored the projected poor performance of SMRs in electricity markets, driven by higher marginal costs for SMRs. Previously, this work presented some evidence that SMR manufacturers may have executed a tradeoff for lower investment costs by increasing operating and marginal costs. This section seeks to observe whether that tradeoff exists, and quantify it under various economic conditions. Considering that investor reticence for traditional nuclear has been driven by escalation of investment costs \cite{locatelli_2018_why_megaprojects}, this section also quantifies the impact of reducing SMR marginal costs on the ceiling for investment costs.

A generic 300MWe SMR was simulated under the best performing price scenarios, the CAISO Historical and Mid Case 100, and tradeoff curves representing the highest possible investment cost at each marginal cost level were built for a given payback period, as illustrated in Figure \ref{fig:figures_of_merit}. Curves were also calculated with both high policy support and high capacity revenues. PTC at $\frac{\$33}{MWh}$ \cite{inlatbinput} and lifetime average capacity prices of $\frac{\$10.14}{kW-month}$ were used as benchmarks for high price and policy support. Two payback period standards, 40 and 20 years, were then used to explore private investment feasibility. Table \ref{tab:baseload_v_current_costsets} outlines the largest investment cost that can be incurred under various scenarios at baseload and current SMR average marginal costs extracted from tradeoff curves.

Results show the clear existence of a tradeoff between marginal and investment costs under all scenarios, as a negative correlation is observed. This observation is significant as it emphasizes a key result of this work: reduced investment costs do not necessarily lead to more economic outcomes, if they come at the expense of higher marginal costs. Observing Figure \ref{fig:figures_of_merit} for how current concepts perform under this paradigm, this work finds that under current investment and marginal cost structures, the majority of SMRs expect to pay back in 40 years only with an expectation of a high policy and price environment. 

Further strengthening earlier findings of this work on marginal costs can be found by observing the slopes of the tradeoff curves in Figure \ref{fig:figures_of_merit}. Steeper curves in the 20 year compared to the 40 year payback period plot speak to the dominance of marginal costs as the target payback period is reduced. For SMRs to be more feasible for private investment, improvements in marginal costs pay larger dividends than investment cost reductions. On average across the price scenarios, if SMRs were to improve marginal costs from current levels of $\frac{\$25}{MWh}$ to baseload costs of $\frac{\$12}{MWh}$ \cite{NREL2024ATB}, investment cost ceilings can be relaxed by 30\%.

Historically high wholesale and capacity prices, with high policy subsidies enable a maximum investment cost of $\frac{\$7,813}{kWe}$ at baseload marginal costs and $\frac{\$6,259}{kWe}$ at current average SMR marginal costs. While current investment costs of SMRs are unclear, if NuScale's early NOAK benchmarks of $\frac{\$20,139}{kWe} $\cite{Schlissel2023_IEEFA_NuScaleCostEstimates} are any indication, it is likely that SMRs would have to achieve steep reductions, at best around 60\% if both marginal cost reductions as well as high policy and price environment. Most likely, SMRs may have to achieve investment cost reductions at a rate similar to Utility PV over the past two decades, greater than 80\% \cite{ramasamy2025pv_costs}, in order to be feasible for private investment.

\subsection{Investment cost escalation and comparison with traditional nuclear}
\label{sec::sensitivities}

Private investment feasibility of nuclear power has been traditionally been highly sensitive to their investment costs. While this work has shown marginal costs as the primary driver behind poor projected performance for SMRs, nuclear power historically has experienced investment cost escalations of 117\% on 97\% of all projects \cite{sovacool2014}. 

\begin{figure*}[p!]
    \centering

    \includegraphics[width=0.65\textwidth]{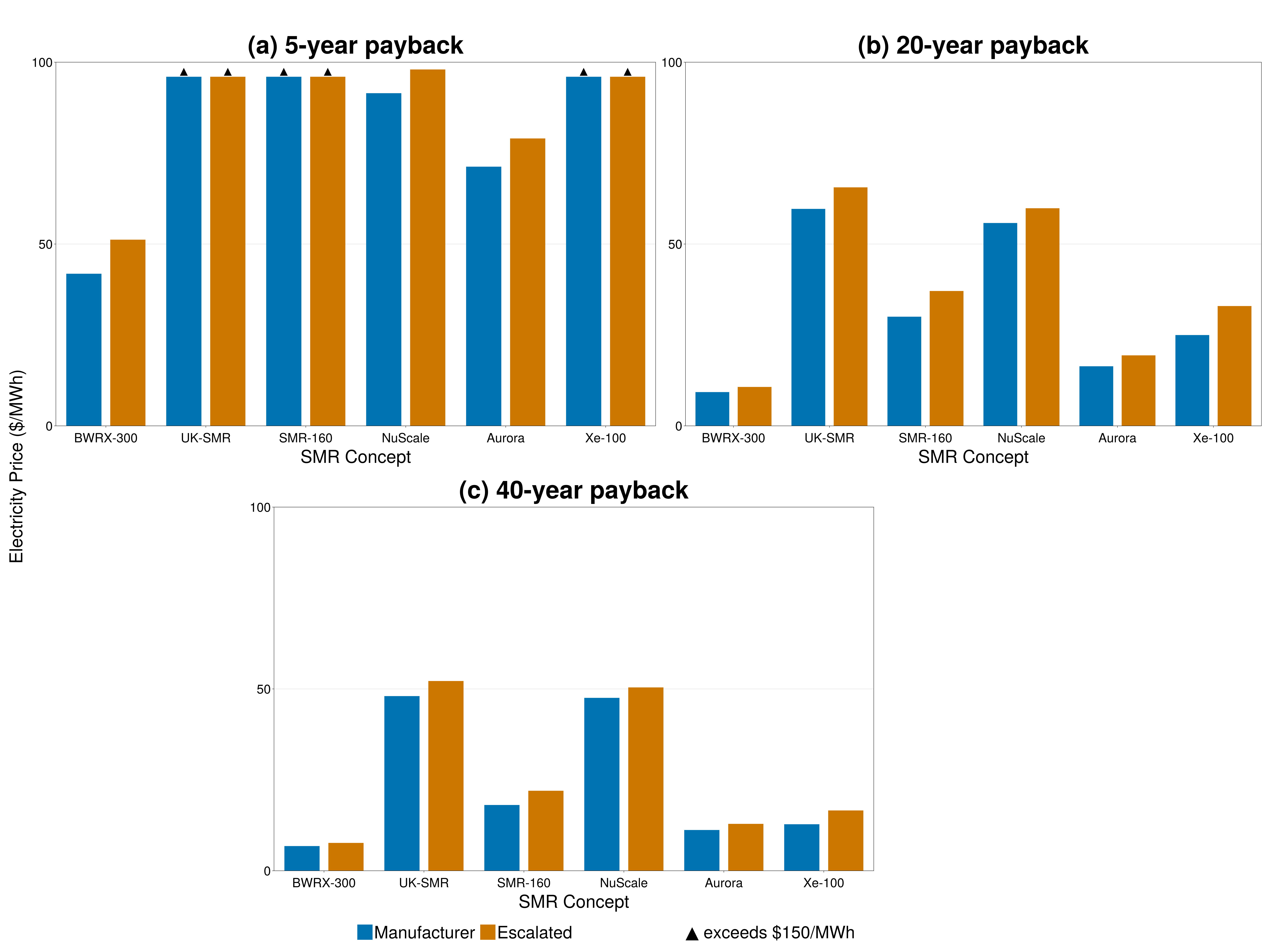}
    \caption{Effects of cost escalation on the average lifetime electricity price required to achieve target payback periods of 5, 20, and 40 years. Missing bars indicate that prices must exceed $\$100/\mathrm{MWh}$ to achieve the target payback period. With a 117\% increase in investment costs, average electricity prices necessary to achieve sub-20-year payback periods increase significantly.}
    \label{fig:costoverrun}

    \vspace{1.5em}

    \includegraphics[width=\textwidth]{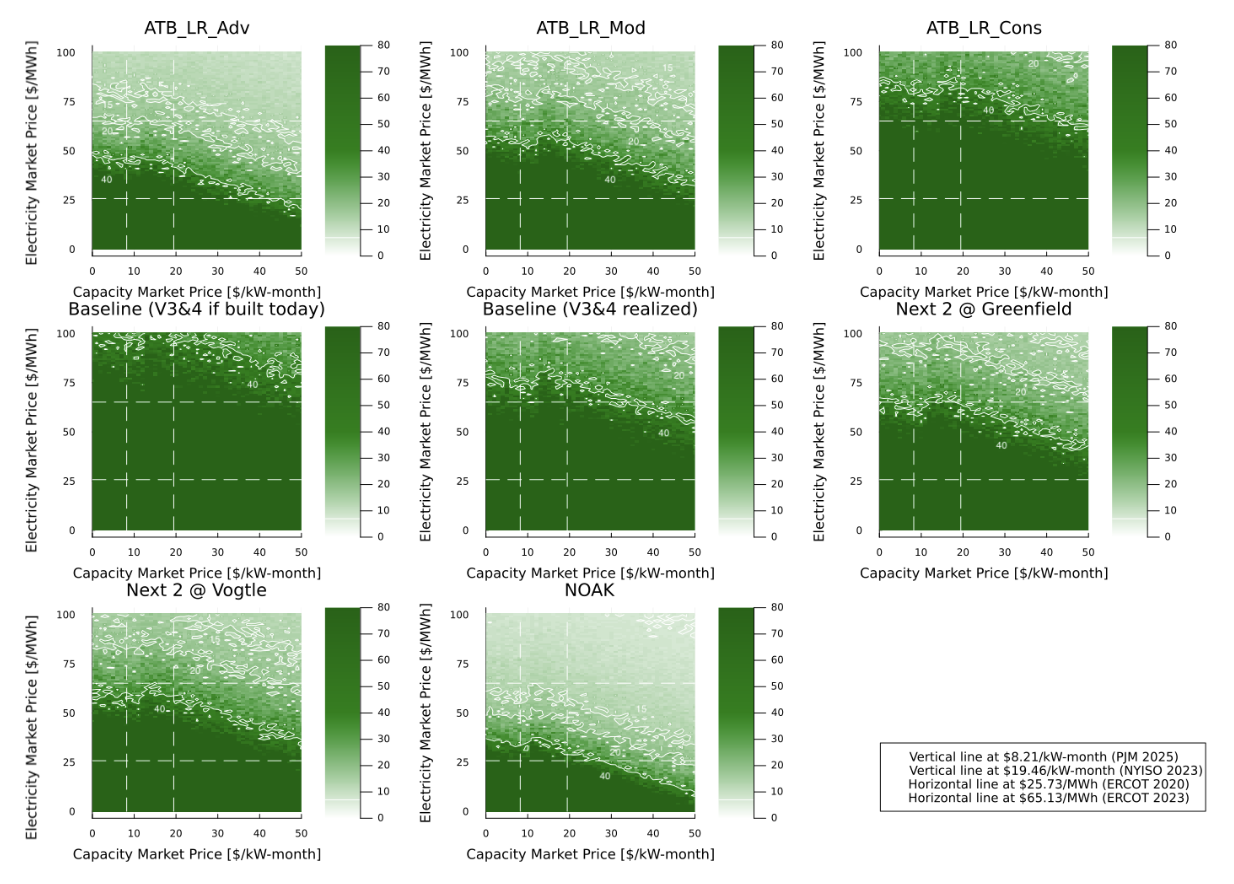}
    \caption{Payback periods of AP1000 cost scenarios as a function of electricity and capacity prices. SMR economic performance at escalated investment costs is similar to the AP1000.}
    \label{fig:ap1000-heatmap}

\end{figure*}

A similar increase in investment costs for SMRs, as seen in Figure \ref{fig:costoverrun}, increases the average lifetime electricity prices required to achieve sub-20 year payback periods on average by $\frac{\$25}{MWh}$, resulting in almost all SMRs being unable to achieve the investment feasibility standard. Even optimistic concepts such as OKLO's Aurora-15 and GE-Hitachi's BWRX-300 hover on the edge of investment feasibility under escalated costs. 

How SMRs compare to traditional nuclear reactors, especially to Georgia Power's Vogtle 3 \& 4 units that entered operation in 2024, is used as a benchmark of whether the economic performance of nuclear power has improved in the transition to SMRs. Building Vogtle resulted in significant cost and schedule overruns and the bankruptcy of its manufacturer (Westinghouse) \cite{vogtlecosts}. SMRs would need to significantly improve on the experience of Vogtle to achieve private investment feasibility.

Realized Vogtle costs and projected future AP1000 cost sets were simulated to compare performance under energy and capacity market structures, seen in Figure \ref{fig:ap1000-heatmap}. This work's final finding is that if SMRs experience investment cost escalations, at a rate outlined in Figure \ref{fig:costoverrun}, the majority of the concepts will have a performance profile similar to moderate and conservative estimates of the Vogtle 3 \& 4, as well as the projected future builds of the AP1000. 

Additionally, NOAK and advanced estimates of the AP1000 perform similarly to most SMRs at manufacturer estimates. If the SMR manufacturer estimates are NOAK too, the perceived benefits of the transition to SMRs are brought into question. While SMRs are less capital intensive, how private investment feasibility would be impacted if the concepts had similar economic performance as traditional reactors is unclear.
\clearpage

\clearpage

\section{Conclusion and Policy Implications}

The results of this work demonstrate that operational cost escalation of SMRs drive their poor project performance in ISO electricity markets, with and without policy support. In an effort to reduce upfront investment cost for SMRs, design tradeoffs such as using novel fuels that do not have established supply chains most likely ballooned operational and marginal costs. As a result, intra-hour losses from operating the SMRs overshadow any upfront investment cost advantages. 

As a consequence, SMRs require an environment of ISO market prices and public subsidies beyond historic highs to attract private investment at manufacturer advertised cost benchmarks, with almost all concepts unable to achieve discounted payback periods less than their technical lifetimes. Average lifetime capacity prices need to exceed PJM's historic high 2025/26 auction price, which caused the Commonwealth of Pennsylvania to threaten to leave PJM \cite{Howland2025_PJMStatesGovernance}. Policy support for investment costs are generally shown to be ineffective, where for some SMRs, even if the ITC subsidized the entire cost of construction it is still insufficient to improve payback periods below technical lifetimes. However, if policy makers were to raise the PTC to $\frac{\$95}{MWh}$, to better reflect the social cost of carbon \cite{socialcostofcarbon} of avoided emissions (primarily natural gas at roughly 0.5 tons/MWh), most SMRs would be able to achieve payback periods below 20 years, with or without capacity revenues. 

If cost reductions were to be made, improvements in marginal costs pay larger dividends than investment cost reductions. If SMR marginal costs improved to baseload traditional nuclear levels at $\frac{\$12}{MWh}$, investment cost benchmarks can be relaxed by an average of 30\%, with a maximum investment cost of $\frac{\$7,813}{kWe}$ under high wholesale and capacity prices, supplemented with high PTC rates. 

Comparing performance to traditional nuclear reactors, this work finds that SMRs at manufacturer estimates are as profitable as estimates of Westinghouse's AP1000. If manufacturer investment cost estimates were to increase at the average historical nuclear power plant rate, SMRs would have similar profitability to Vogtle 3 \& 4. This sharply brings into question the benefits of the transition to advanced nuclear plants, especially since traditional plants enjoy economies of scale that SMRs lack.

Future work to explore the operational value of nuclear power to a power grid would be beneficial, as well as an examination of performance under global electricity market structures. Additionally, performance under novel proposed market structures in the United States should be explored, especially with a focus on future power grid scenarios.

\section{Acknowledgments}
The authors would like to thank Dr. Charles F. Kutscher for his review and helpful feedback throughout the process of this paper.
\printcredits

\clearpage
\appendix

\section{Additional Reactor Data}

\begin{strip}
\centering
\captionof{table}{Full list of manufacturer-advertised cost data taken from Steigerwald et al. (2023) and expanded by adding additional concepts. Missing variable cost data use Lazard 2024 assumptions \cite{lazard2024lcoe}. Fixed costs have been converted to \$/MWh for comparison purposes.}
\label{tab:smrcostdata}

\scriptsize
\setlength{\tabcolsep}{4pt}
\renewcommand{\arraystretch}{1.08}

\resizebox{\textwidth}{!}{%
\begin{tabular}{lrrrrr}
\toprule
Project & Capacity [MWel] & Investment Cost [\$/MWel] & Fuel Cost [\$/MWh] & Fixed O\&M Cost [\$/MWh] & Variable O\&M Cost [\$/MWh] \\
\midrule
BWRX-300 & 300.0 & 2,250,000 & 8.44 & 16.00 & 3.55 \\
UK-SMR & 443.0 & 5,215,937 & 10.95 & 68.00 & 3.55 \\
SMR-160 & 160.0 & 6,312,500 & 14.48 & 11.83 & 3.55 \\
SMART & 107.0 & 10,186,916 & 4.88 & 7.44 & 3.55 \\
NuScale & 77.0 & 3,466,000 & 17.42 & 72.26 & 3.55 \\
RITM 200M & 53.0 & 4,212,000 & 26.82 & 19.18 & 3.55 \\
ACPR 50S & 40.0 & 8,532,000 & 16.73 & 11.83 & 3.55 \\
KLT-40S & 35.0 & 13,531,429 & 91.18 & 19.18 & 3.55 \\
CAREM & 30.0 & 23,187,500 & 19.24 & 17.13 & 3.55 \\
EM2 & 265.0 & 4,373,300 & 11.97 & 11.91 & 3.55 \\
HTR-PM & 105.0 & 5,400,000 & 15.47 & 18.95 & 3.55 \\
PBMR-400 & 165.0 & 1,550,000 & 8.63 & 16.74 & 3.55 \\
ARC-100 & 100.0 & 5,050,000 & 27.15 & 19.93 & 0.31 \\
CEFR & 20.0 & 23,034,536 & 113.05 & 19.93 & 3.55 \\
4S & 10.0 & 2,500,000 & 48.67 & 38.99 & 17.26 \\
IMSR (300) & 195.0 & 4,054,266 & 8.12 & 18.69 & 3.55 \\
SSR-W & 37.5 & 1,950,000 & 21.63 & 13.85 & 3.55 \\
e-Vinci & 3.5 & 5,771,429 & 2.35 & 22.75 & 3.55 \\
Brest-OD-300 & 300.0 & 4,160,000 & 30.62 & 26.27 & 3.55 \\
Aurora-15 \cite{oklo2023investor} & 15.0 & 3,800,000 & 13.41 & 15.51 & 4.25 \\
Aurora-50 \cite{oklo2023investor} & 50.0 & 2,320,000 & 6.64 & 10.86 & 3.40 \\
Xe-100 \cite{Xenergy_AAC_2022} & 80.0 & 7,500,000 & 3.39 & 2.50 & 3.55 \\
\bottomrule
\end{tabular}%
}
\end{strip}

\vspace{1em}

\begin{strip}
\centering
\captionof{table}{Additional operational data for SMR concepts.}
\label{tab:smropdata}

\scriptsize
\setlength{\tabcolsep}{4pt}
\renewcommand{\arraystretch}{1.08}

\resizebox{\textwidth}{!}{%
\begin{tabular}{lrrrrr}
\toprule
Project & Lifetime [yr] & \# Modules & Construction Duration [mo.] & Refueling Minimum Time [mo.] & Refueling Maximum Time [mo.] \\
\midrule
BWRX-300 & 60 & 1 & 26 & 12 & 24 \\
UK-SMR & 60 & 1 & 24 & 18 & 24 \\
SMR-160 & 80 & 2 & 36 & 20 & 26 \\
SMART & 60 & 1 & 36 & 27 & 33 \\
NuScale & 60 & 12 & 36 & 15 & 18 \\
RITM 200M & 60 & 2 & 48 & 51 & 57 \\
ACPR 50S & 40 & 1 & 30 & 26 & 32 \\
KLT-40S & 40 & 2 & 48 & 30 & 36 \\
CAREM & 40 & 1 & 108 & 12 & 16 \\
EM2 & 60 & 4 & 42 & 15 & 18 \\
HTR-PM & 40 & 2 & 48 & 32 & 36 \\
PBMR-400 & 40 & 1 & 34 & 70 & 74 \\
ARC-100 & 60 & 1 & 54 & 230 & 250 \\
CEFR & 30 & 1 & 120 & 15 & 18 \\
4S & 60 & 1 & 12 & 360 & 360 \\
IMSR (300) & 60 & 2 & 55 & 80 & 88 \\
SSR-W & 60 & 8 & 31 & 80 & 88 \\
e-Vinci & 40 & 2 & 12 & 36 & 96 \\
Brest-OD-300 & 30 & 4 & 60 & 29 & 48 \\
Aurora-15 \cite{oklo2023investor} & 40 & 1 & 12 & 118 & 120 \\
Aurora-50 \cite{oklo2023investor} & 40 & 1 & 12 & 118 & 120 \\
Xe-100 \cite{Xenergy_AAC_2022} & 60 & 4 & 36 & 6 & 12 \\
\bottomrule
\end{tabular}%
}
\end{strip}

\bibliographystyle{cas-model2-names}

\bibliography{cas-refs}



\end{document}